\documentclass[reprint, superscriptaddress, aps, showkeys, amsmath,amssymb]{revtex4-1}

\usepackage{graphicx}
\usepackage{bm}
\usepackage{xfrac}
\usepackage[table]{xcolor}
\usepackage{tikz}
\usetikzlibrary{decorations.pathreplacing, shadows, shapes}
\usepackage[qm]{qcircuit}
\usepackage{multirow} 
\usepackage{setspace}
\usepackage{booktabs}
\usepackage{hyperref}
\usepackage{multirow}
\usepackage{setspace}
\usepackage{booktabs}
\usepackage{hyperref}
\usepackage{url}
\usepackage{siunitx} 

\definecolor{mydarkred}{RGB}{153, 0,0} 
\definecolor{mydarkblue}{RGB}{0,128,128}

\newcommand{\ket}[1]{| #1 \rangle} 

\newcommand{\x}{\mathbf{x}}
\newcolumntype{C}[1]{>{\centering\let\newline\\\arraybackslash\hspace{0pt}}m{#1}}
\newcolumntype{N}{@{}m{0pt}@{}}

\begin{document}

\title{Implementing a distance-based classifier with a quantum interference circuit}

\author{Maria Schuld}
\email[]{schuld@ukzn.ac.za}
\affiliation{Quantum Research Group, School of Chemistry and Physics, University of KwaZulu-Natal, Durban 4000, South Africa}
\author{Mark Fingerhuth}
\email[]{markfingerhuth@protonmail.com}
\affiliation{Quantum Research Group, School of Chemistry and Physics, University of KwaZulu-Natal, Durban 4000, South Africa}
\affiliation{Maastricht Science Programme, University of Maastricht, 6200 MD Maastricht, The Netherlands}
\author{Francesco Petruccione}
\email[]{petruccione@ukzn.ac.za}
\affiliation{Quantum Research Group, School of Chemistry and Physics, University of KwaZulu-Natal, Durban 4000, South Africa} 
\affiliation{National Institute for Theoretical Physics, KwaZulu-Natal, Durban 4000, South Africa}

\begin{abstract}
Lately, much attention has been given to quantum algorithms that solve pattern recognition tasks in machine learning. Many of these quantum machine learning algorithms try to implement classical models on large-scale universal quantum computers that have access to non-trivial subroutines such as Hamiltonian simulation, amplitude amplification and phase estimation. We approach the problem from the opposite direction and analyse a distance-based classifier that is realised by a simple quantum interference circuit. After state preparation, the circuit only consists of a Hadamard gate as well as two single-qubit measurements, and computes the distance between data points in quantum parallel. We demonstrate the proof-of-principle using the IBM Quantum Experience and analyse the performance of the classifier with numerical simulations, showing that it classifies surprisingly well on simple benchmark tasks.  
\end{abstract}

\pacs{03.67.Ac,07.05.Mh}

\keywords{Quantum machine learning, quantum algorithm, supervised classification, state preparation, IBM Quantum Experience}

\maketitle

Quantum machine learning, an emerging discipline combining quantum computing with intelligent data mining, witnesses a growing number of proposals for quantum algorithms that solve the problem of supervised pattern recognition \cite{rebentrost14,kapoor16, benedetti16b,denchev12,schuld16lr}. In supervised learning, a dataset of labelled inputs or feature vectors is given, and the task is to predict the label of a new feature vector. For example, the inputs could be images of persons, while the label is the name of the person shown in the picture. Image recognition software is then supposed to recognise which person is shown in a previously unseen image. A central question of quantum machine learning is if and how a quantum computer could enhance methods known from machine learning \cite{dunjko16}. A large share of the suggested quantum machine learning algorithms are based on the assumption of a large-scale universal quantum computer that can implement nontrivial circuits. This is specifically true for distance-based machine learning models: Quantum algorithms for $k$-nearest neighbour and clustering have been based on extensions of amplitude amplification \cite{wiebe14,aimeur07} while quantum kernel methods such as support vector machines \cite{rebentrost14} and Gaussian processes \cite{zhao15} rely on the rather technical  routines for quantum matrix inversion \cite{harrow09} or density matrix exponentiation \cite{lloyd14}. Experimental implementations are necessarily limited to demonstrations that concede a vast reduction in complexity of the quantum circuits \cite{cai15,li15}. Most importantly, quantum machine learning remains an enterprise to merely mimic methods from classical machine learning that have been tailor-made for classical computation. \\ 
The aim of this Letter is to propose a change in perspective: We start with the most simple quantum circuit and show that it can be used as a -- likewise simple -- model of a classifier. Instead of choosing a textbook machine learning algorithm and asking how to run it on a quantum computer, we turn the question around and ask what classifier can be realised by a minimum quantum circuit. The basic idea is to use quantum interference to evaluate the distance measure of a kernel classifier in quantum parallel. A similar idea has been investigated by some of the authors in ref.\cite{schuld14neigh}, but based on a less powerful information encoding strategy and a more complex circuit. \\
If an efficient state preparation routine is known, the algorithm explored here harvests the same logarithmic scaling in the dimension and number of the input data that has been claimed by other authors \cite{zhao15,rebentrost14}, but only requires a relatively simple setup that can easily be implemented on small-scale quantum computing devices available today \cite{mohseni17}. Evidently, by using only a single-qubit gate this ``speedup'' is not necessarily based on quantum resources. However, besides the argument we want to make we envision this to be interesting in situations where quantum states generated by quantum simulations - for example in quantum chemistry - have to be classified coherently. This case is sometimes referred to as `quantum data'. \\
In order to demonstrate the circuit, a simplified supervised pattern recognition task based on the famous Iris flower dataset is solved with the 5-qubit quantum computer provided by the IBM Quantum Experience \cite{ibmquantumcomputer}. Since at the time of writing the interface only allowed an implementation of $80$ quantum gates, numerical simulations show that the classifier performs well enough in simple benchmark tasks.\\

We consider the task of supervised binary pattern classification which can be formalised as follows: Given a training dataset $\mathcal{D} = \{(\x^1,y^1),...,(\x^M, y^M)\}$ of inputs $\x^m \in \mathbb{R}^N$ with their respective target labels $y^m \in \{-1,1\}$ for $m=1,...,M$, as well as a new unlabeled input $\tilde{\x} \in \mathbb{R}^N$, find the label $\tilde{y} \in \{-1,1\}$ that corresponds to the new input. The classifier effectively implemented by the quantum interference circuit together with a thresholding function is given by:
\begin{equation}
\tilde{y} = \mathrm{sgn} \left(\sum\limits_{m=1}^M y^m \Big[ 1- \frac{1}{4M}|\tilde{\x} - \x^m|^2\Big] \right).
\label{Eq:classifier}
\end{equation}
The distance measure $ \kappa(\x, \x') = 1- \frac{1}{4M}|\x - \x'|^2$  can be interpreted as a kernel (and is in fact very similar to an Epanechnikov kernel \cite{epanechnikov69}). The model in eq. (\ref{Eq:classifier}) therefore has the standard form of a kernelised binary classifier, $\tilde{y} = \mathrm{sgn} \left( \sum_{m}w_m y^m \kappa(\tilde{\x}, \x^m)  \right)$ with uniform weights $w_m =1$ \cite{aizerman64}. Such a model can be derived from a perceptron in which the original weights are expressed by an expansion of the training data as motivated by the \textit{representer theorem} \cite{scholkopf01}, and inner products between inputs are replaced by another kernel function via the ``kernel trick''. The model relates to $k$-nearest-neighbour when setting $k\rightarrow M$ and weighing the neighbours by the distance measure.\\ 

\begin{figure}
    \centering
      \includegraphics[width=0.3\textwidth]{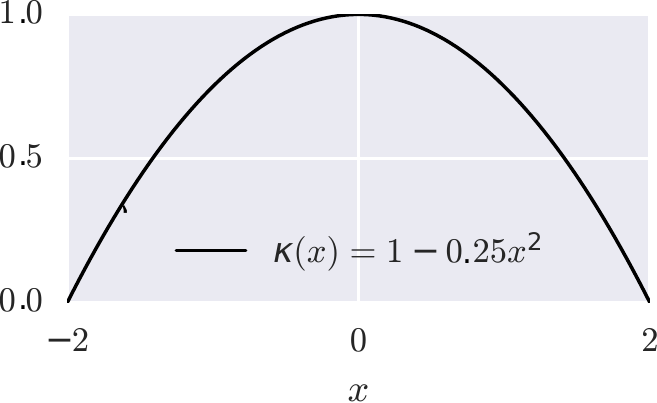}   
\caption{Plot of the kernel function $\kappa(\x, \x') = 1- \frac{1}{4M}|\x - \x'|^2$ on the interval $[-2,2]$ and for $M=1$. The kernel defines the distance measure of the binary classifier considered in this paper. For unit length feature vectors, the maximum distance between two vectors is $2$, and the kernel does not take negative values.}
     \label{Fig:kernel}
\end{figure}
The quantum machine learning algorithm that implements the classifier from eq. (\ref{Eq:classifier}) is based on the idea to encode the input features into the amplitudes of a quantum system and manipulate them through quantum gates - a strategy responsible for most claims of exponential speedups in quantum-enhanced machine learning. We will refer to this approach as `amplitude encoding' to distinguish it from the more common practice of encoding one bit of information into a qubit. Given a classical vector $\x \in \mathbb{R}^N$, where without loss of generality $N$ is assumed to be the $n$th power of two, $N = 2^n$ (which can be achieved by padding the vector with zero entries). Furthermore, assume that $\x$ is normalised to unit length, $\x^T\x = 1$. Amplitude encoding associates $\x = (x_1,...,x_N)^T$ with the $2^n$ amplitudes describing the state of a $n$-qubit quantum system, $\ket{\psi_{\x}} = \sum_{i=0}^{N-1} x_i \ket{i}$.
Here, $\ket{i}$ is an \textit{index register} that flags the $i$th entry of the classical vector with the $i$th computational basis state.

If one can find an efficient quantum algorithm (i.e., with resources growing polynomially in the number of qubits $n$), one manipulates the $2^n$ amplitudes `super-efficiently' (i.e., with resources growing logarithmically in the dimension of the Hilbert space, $\mathcal{O}(\log N)$).  A `super-efficient' algorithm can only maintain its speed if data encoding into a quantum state is also at most polynomial in the number of qubits. There are cases for which this is known to be possible \cite{grover02,soklakov06}. A proposal frequently referred to is a Quantum Random Access Memory \cite{giovannetti08,rebentrost14,zhao15} that loads the bit strings representing $x_i$ in parallel into a qubit register and performs a conditional rotation and measurement of an ancilla to write the values into the amplitude.\\

The chance of success of this postselective measurement is only high if the $x_i$ are uniformly close to one.\\
Using a suitable state preparation scheme, the quantum classification circuit takes a quantum system of $n$ qubits in state
\begin{equation}
\ket{\mathcal{D}} = \frac{1}{\sqrt{2 M C}} \sum_{m=1}^M \ket{m} \Big( \ket{0} \ket{\psi_{\tilde \x}} + \ket{1} \ket{\psi_{\x^m}} \Big) \ket{y^m} . 
\label{Eq:initial_state}
\end{equation}
Here,  $\ket{m}$ is an index register running from $m=1,...,M$ and flagging the $m$th training input. The second register is a single ancilla qubit whose ground state is entangled with the third register encoding the $m$th training state, $\ket{\psi_{\x^m}} = \sum_{i=0}^{N-1} x^m_i\ket{i}$, while the excited state is entangled with the third register encoding the new input $\ket{\psi_{\tilde \x}} = \sum_{i=0}^{N-1} \tilde{x}_i \ket{i}$. The fourth register is a single qubit, which is zero if $y^m = -1$ and one if $y^m =1$. Effectively, this creates an amplitude vector which contains the training inputs as well as $M$ copies of the new input. The normalisation constant $C$ depends on the preprocessing of the data. We will assume in the following that the feature vectors are normalised and hence $C=1$.  \\
After state preparation, the quantum circuit only consists of three operations. First, a Hadamard gate on the ancilla interferes the copies of the new input and the training inputs,
$$ \frac{1}{2\sqrt{ M}} \sum_{m=1}^M  \ket{m} \Big( \ket{0} \ket{\psi_{\tilde \x + \x^m}} \big) +\ket{1}\big( \ket{\psi_{\tilde \x - \x^m}} \big)  \Big) \ket{y^m} , 
$$
where $\ket{\psi_{\tilde \x \pm \x^m}} = \ket{\psi_{\tilde \x}} \pm \ket{\psi_{\x^m}}$.
The second operation is a conditional measurement selecting the branch with the ancilla in state $\ket{0}$. This postselection succeeds with probability $\mathrm{p}_{\mathrm{acc}} = \frac{1}{4M} \sum _m  |\tilde{\x} + \x^{m}|^2$. It is more likely to succeed if the collective Euclidean distance of the training set to the new input is small. We will show below that if the data is standarised, postselection usually succeeds with a probability of around $0.5$. 
If the conditional measurement is successful, the result is given by
$$ \frac{1}{2\sqrt{ M \mathrm{p}_{\mathrm{acc}}}} \sum\limits_{m=1}^M \sum\limits_{i=1}^{N} \ket{m} \left( \tilde{x}_i + x^{m}_i\right) \ket{i}  \ket{y^{m}} .$$
The amplitudes weigh the class qubit $\ket{y^m}$ by the distance of the $m$th data point to the new input. In this state, the probability of measuring the class qubit $\ket{y^{m}}$ in state $0$,
$$\mathrm{p}(\tilde{y} = 0)= \frac{1}{4 M \mathrm{p}_{\mathrm{acc}}} \sum\limits_{m| y^m = 0}  | \tilde{\x} + \x^{m} |^2 ,$$
reflects the probability of predicting class $-1$ for the new input. The choice of normalised feature vectors ensures that$\frac{1}{4 M \mathrm{p}_{\mathrm{acc}}}\sum _m |\tilde{\x} + \x^{m} |^2= 1-\frac{1}{4 M \mathrm{p}_{\mathrm{acc}}}\sum _m |\tilde{\x} - \x^{m}|^2$, and choosing the class with the higher probability therefore implements the classifier from eq. (\ref{Eq:classifier}). The Supplementary Material shows that the number of measurements needed to estimate $\mathrm{p}(\tilde{y} = 0)$ to error $\epsilon$ with a reasonably high confidence interval grows with $\mathcal{O}(\epsilon^{-1})$.  \\
As a demonstration we implement the interference circuit with the IBM Quantum Experience  (IBMQE) \cite{ibmquantumcomputer} using the Iris dataset \cite{fisher36}. Data preprocessing consists of two steps (see fig. \ref{Fig:preproc}): We first standardise the dataset to have zero mean and unit variance. This is common practice in machine learning to compensate scaling effects, and in our case ensures that the data does not only populate a small subspace of the input space, which in higher dimensions leads to indistinguishably small distances between data points. Second, we need to normalise each feature vector to unit length. This strategy is popular in machine learning - for example with support vector machines - to only consider the angle between data points. (As an intuition, if we want to classify flowers, some items may have grown bigger than others due to better local conditions, but it is the proportion of the sepal and petal length that is important for the class distinction). This preprocessing strategy allows us to fulfill the conditions of `super-efficient' preprocessing in refs. \cite{soklakov06,giovannetti08} 
\begin{figure}
\centering
\includegraphics[width=0.45\textwidth]{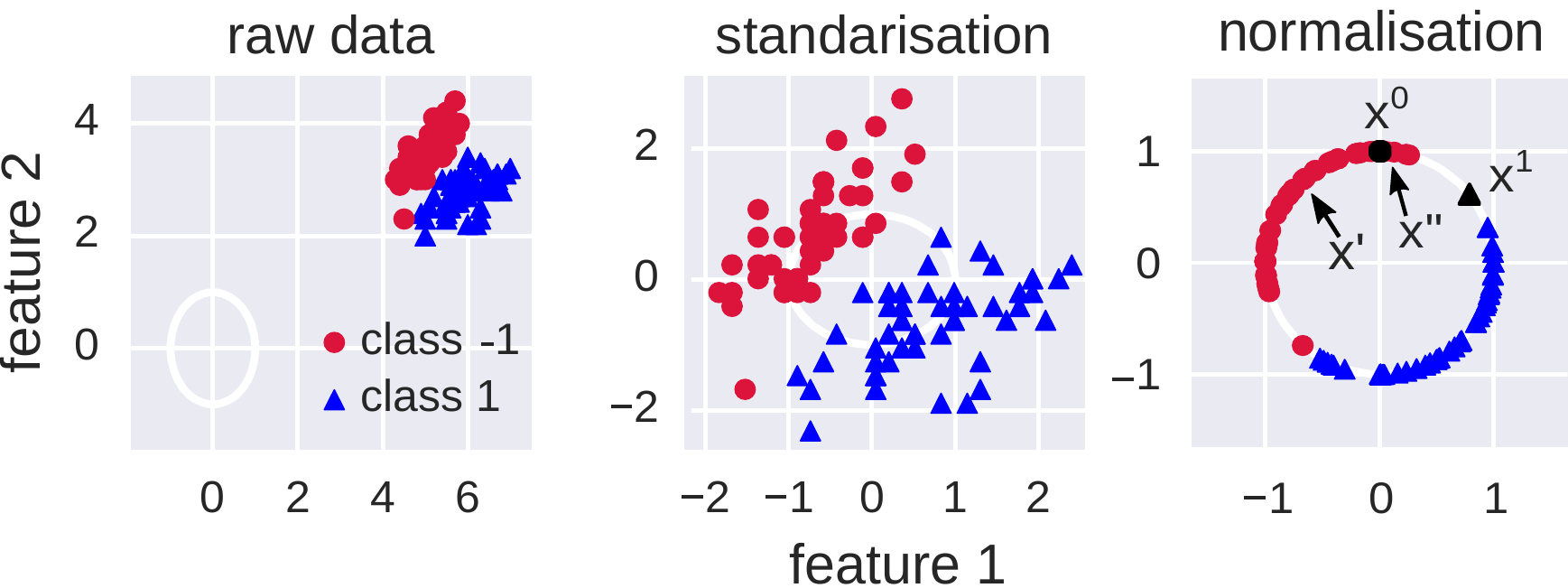}   
\caption{Data processing illustrated with the example of the first two classes (here called $-1$ and $1$) of the first two features of the Iris dataset. The raw data (left) gets standarised to zero mean and unit variance (center), after which each feature vector is normalised to unit length  (right). The training points used in the experiment are marked in black, while the arrows point to the new feature vectors to classify.}
\label{Fig:preproc}
\end{figure}
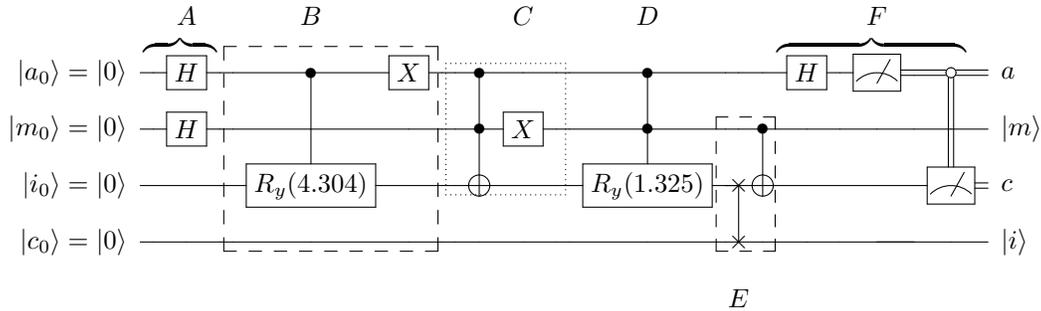
\begin{figure*}

%
%
$$
\Qcircuit @C=.5em @R=0.5em @!R {
& & & & & & &  & A &  &  & B &  & &  &  & C &  &  & D &  & &  &  &  & F &  & & \\
& & & & & &\lstick{\ket{a_0}=\ket{0}} & \qw & \gate{H} & \qw & \qw & \ctrl{2} & \gate{X} & \qw & \qw & \ctrl{1} & \qw & \qw & \qw & \ctrl{1} & \qw & \qw & \qw & \gate{H} & \qw & \meter & \cw &  \cctrlo{2} &\rstick{a} \cw \\
& & & & & & \lstick{\ket{m_0}=\ket{0}} & \qw & \gate{H} & \qw & \qw  & \qw & \qw & \qw & \qw & \ctrl{1} & \gate{X} & \qw & \qw & \ctrl{1} & \qw & \qw & \ctrl{1} & \qw  & \qw  & \qw & \qw & \qw  &\rstick{\ket{m}} \qw \\
& & & & & & \lstick{\ket{i_0}=\ket{0}} & \qw & \qw & \qw & \qw & \gate{R_y(4.304)}  & \qw & \qw & \qw & \targ & \qw & \qw & \qw & \gate{R_y(1.325)} & \qw & \qswap \qw & \targ & \qw  & \qw  & \qw & \qw & \meter & \rstick{c} \cw\\
& & & & & & \lstick{\ket{c_0}=\ket{0}} & \qw & \qw & \qw & \qw  & \qw & \qw & \qw & \qw & \qw & \qw & \qw & \qw & \qw & \qw & \qswap \qwx & \qw & \qw & \qw & \qw & \qw\qw & \qw & \rstick{\ket{i}} \qw \gategroup{2}{8}{4}{9}{.7em}{^\}}{A}
\gategroup{2}{11}{5}{13}{.7em}{--} \gategroup{2}{15}{4}{18}{.7em}{.} \gategroup{2}{24}{5}{28}{.7em}{^\}} \gategroup{3}{21}{5}{23}{.7em}{--}\\
& & & & & & &  &  &  &  &  & & &  &  &  &  &  & &  & E &  &  &  &  &  & & \\
} 
$$
\caption{\label{Fig:2D_full_circuit} Quantum circuit implementing the distance-based classifier using the two training vectors $\x^0$ and $\x^1$ and the input vector $\tilde \x'$ from the rescaled and normalised Iris flower dataset. First the ancilla and index qubits are put into uniform superposition (step A) and the input vector $\tilde \x'$  is entangled with the ground state of the ancilla (step B). Then the training vector $\x^0$ is entangled with the excited state of the ancilla and the ground state of the index qubit (step C) followed by entangling training vector $\x^1$ with the excited state of the ancilla and the index qubit (step D). Next, the data and class qubits are swapped and the class qubit is flipped conditioned on the index qubit being $\ket{1}$ (step E) which completes the initial state preparation. In step F, the Hadamard gate interferes the copies of $\tilde \x'$ with the training vectors and the ancilla is measured followed by a measurement of the class qubit (due to prior swapping now at the position of the $
\ket{i}$ qubit) when the ancilla was found to be in the $\ket{0}$ state.}
\end{figure*}
The IBM Quantum Experience enables public use of a processor of five non-error-corrected superconducting qubits based on Josephson junctions located at the IBM Quantum Lab at the Thomas J Watson Research Center in Yorktown Heights, New York. The current processor has limited connectivity between the five qubits and allows the implementation of $80$ gates from a set of $12$ single-qubit quantum logic gates as well as a CNOT gate  (see Supplementary Material for details). Due to these limitations, we will only use the first two features of two samples from the Iris dataset for the experimental implementation of the quantum algorithm. Consider the preprocessed training dataset $\mathcal{D}_1 = \{(\x^0, y^0),(\x^1, y^1) \}$ with the two training vectors $\x^0 = (0,1)$, $y^0 = -1$ (Iris sample 33) and $\x^1 = (0.789, 0.615)$, $y^1 = 1$ (Iris sample 85). In two separate experiments we will consider the classification of two new input vectors of class $-1$ but with varying distances to the training points, $\tilde \x' = (-0.549, 0.836)$ (Iris sample 28) and $\tilde \x'' = (0.053 , 0.999)$ (Iris sample 36) (see fig. \ref{Fig:preproc}).

\begin{figure}
\centering
\includegraphics[width=0.4\textwidth]{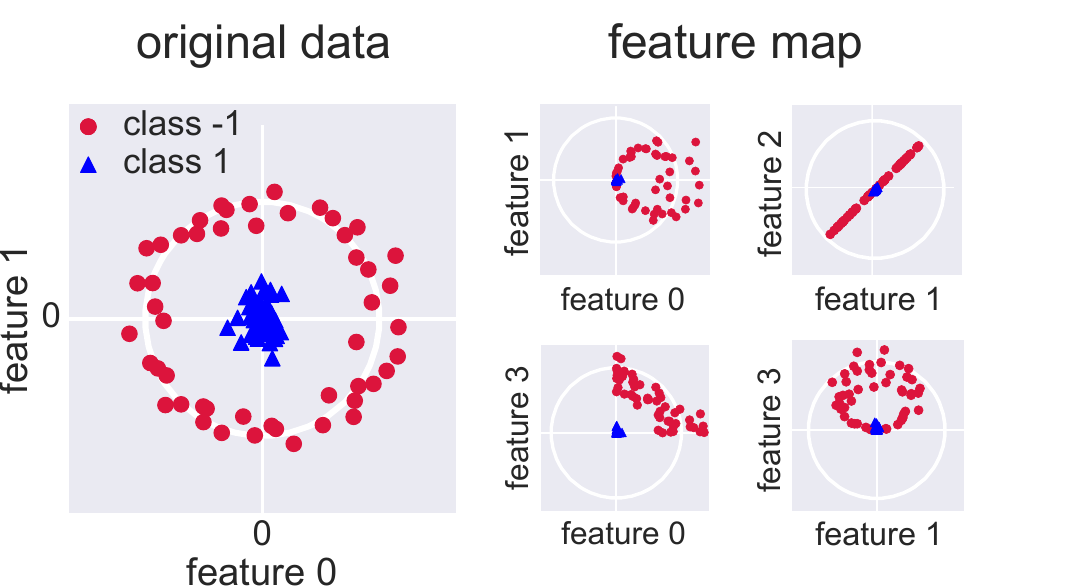}   
\caption{The ``circles'' dataset is not suitable for a distanced-based classifier. However, by using two copies of the quantum states that represent feature vectors one can implement polynomial feature maps that increase the power of the classifier. The four plots on the right display the dataset after the feature map, standarisation and normalisation.}
\label{Fig:kerneltrick}
\end{figure}

Implementing this particular classification problem requires four qubits;  one qubit for the index register $\ket{m}$ to represent two training vectors, one ancilla qubit, one qubit storing the class of each training instance and one qubit for the data register $\ket{i}$ to represent the two entries of each training and input vector as
\begin{eqnarray}
\ket{\psi_{\tilde \x'}} &=& -0.549 \, \ket{0} + 0.836 \, \ket{1} , \\ \nonumber
\ket{\psi_{\tilde \x''}} &=& 0.053 \, \ket{0} + 0.999 \, \ket{1} , \\ \nonumber
\ket{\psi_{\x^0}} &=& \ket{1} ,\\ \nonumber
\ket{\psi_{\x^1}} &=& 0.789 \, \ket{0} + 0.615 \, \ket{1} . \nonumber
\end{eqnarray}
In this small-scale example efficient state preparation does not require sophisticated routines as discussed above, but can be designed by hand (see fig. \ref{Fig:2D_full_circuit}). The main idea is to use controlled rotation gates such that the input and training vectors become entangled with the corresponding states of the ancilla and index qubits. Two aspects have to be considered in the quantum circuit design. Firstly, the single and double controlled rotation gates (step B and D in fig. \ref{Fig:2D_full_circuit}) as well as the Toffoli gate (see step C in fig. \ref{Fig:2D_full_circuit}) required for the entanglement of the ancilla and index qubit with the training vectors $\x^0$ and $\x^1$ are not part of IBM's universal gate set.
Therefore, the state preparation routine needs to be mapped to the available hardware by decomposing the controlled rotation, Toffoli and SWAP gates (see Supplementary Material). Secondly, state preparation for this classification problem requires at least one CNOT operation between qubits that are not directly connected in the hardware.  This problem can be solved by exchanging adjacent qubits with a SWAP gate such that the CNOT operation between previously unconnected qubits becomes feasible (see step E in fig. \ref{Fig:2D_full_circuit}). \\
Using the IBMQE, the resulting quantum circuits were first simulated in an error-free environment and then executed on the non error-corrected hardware for the maximum number of $8192$ runs, and the results are summarised in Table \ref{tab:2D_case_results}. 
As expected the quantum circuits yield simulation results that closely resemble the theoretical predictions, while the experimental results show substantial errors.
The main reasons for these deviations are the lack of error correction, the ancilla qubit's short transversal coherence time causing it to deviate from the initial uniform superposition, as well as the class qubit's short longitudinal coherence time. This mostly affects the class qubit in state $\ket{1}$ storing the class of training vector $\x^1$ which will unavoidably decohere to its ground state.
The rapid decoherence of the class qubit is especially troublesome since it makes classification of input vectors that are expected to be classified as class $\tilde{y} = 1$ impossible and, thus, only input vectors of class $\tilde{y} = -1$ are presented in this Letter. In those cases, despite the large deviation between theory and experiment, both input vectors $\tilde \x'$ and $\tilde \x''$ were correctly classified as class $\tilde{y} = -1$. 
This example demonstrates how mapping a quantum machine learning algorithm to the available hardware requires adaptations to fit the device architecture, and how state preparation, even in this simplified example, constitutes the main bottleneck in the execution of many quantum machine learning algorithms.\\
\begin{table}
\def\arraystretch{1.2}
\begin{tabular}{ C{1.1cm}   C{1.5cm}  C{1.8cm} C{1.8cm}  }
      \hline
      \hline
      Input vector & $\mathrm{p}_\mathrm{acc}$ & $\mathrm{p}(\ket{c} = \ket{0})$ & $\mathrm{p}(\ket{c} = \ket{1})$  \\
            \hline
      &0.455 &0.516&0.484\\
      $\tilde \x'$ & 0.731$^\triangleright$&0.629$^\triangleright$ &0.371$^\triangleright$\\ 
      &0.729*&0.629* &0.371*\\\hline
      &0.494&0.589 &0.411\\
      $\tilde \x''$ &0.911$^\triangleright$ &0.548$^\triangleright$&0.452$^\triangleright$ \\
      &0.913*&0.547*& 0.453*\\
      \hline \hline
    \end{tabular}
    
    \caption{\label{tab:2D_case_results} Classification results for the two-dimensional input vectors $\tilde \x'$ and $\tilde \x''$ from the Iris flower dataset. Experimental results are always shown on top of their corresponding simulation result (marked with triangle) and theoretical prediction (marked with asterisks).}
\end{table}
\begin{table}
\def\arraystretch{1.2}
\begin{tabular}{l  l c c c}
\hline \hline
Dataset & test error& variance & $\mathrm{p}_\mathrm{acc}$\\\hline
Iris class 1\&2 &$0.00$ &$0.000$ & $0.50$ \\ 
Iris class 1\&3 &$0.00$ &$0.000$ & $0.50$ \\ 
Iris class 2\&3 &$0.07$ &$0.003$ & $0.50$ \\ 
Iris class 2\&3, feat map  &   $0.00$ &$0.000$& $0.50$\\
Circles & $0.62$ & $0.006$ & $0.50$ \\
Circles, feat map  &   $0.00$ &$0.000$& $0.55$\\
\hline \hline
\end{tabular} 
\caption{Test error for the quantum classifier on different datasets for $1000$ random separations into test and training set. Using feature maps leads to a zero classification error in the examples of classes 2\& 3 of the Iris data set as well as the circles data set. }
\label{Tbl:sim_results}
\end{table}
In order to analyse the performance of the classifier in more depth, we finally present some numerical simulations. The Iris dataset was randomly divided into a training and test set (ratio $80/20$). Table \ref{Tbl:sim_results} shows the error, or the proportion of misclassified test instances to all test instances, for $1000$ repetitions of the random division. The variance of the error is very low in all cases, and the acceptance probability of selecting the correct branch is around $0.5$ due to the standarisation of the data. The results show that running the quantum classifier circuit on the entire dataset for classes $1$ and $2$ as well as  for $1$ and $3$ of Iris species results in a $100\%$ success rate while distinguishing classes $2$ and $3$ only yield a $93\%$ average success rate. This is due to the fact that classes $2$ and $3$ are not linearly separable and cannot be distinguished easily by distance-based methods - especially with a broad kernel that considers almost the entire input space. However, working with normalised feature vectors makes the classifier amenable for the ``kernel trick'', in which replacing the inner product in the distance by another kernel effectively implements a map on the input space which can vastly improve the power of the classifier \cite{scholkopf00}. This has also been proposed for quantum machine learning \cite{rebentrost14,chatterjee16}. Implementing a polynomial feature map \cite{rebentrost14} requires to prepare copies of the quantum states which represent the feature vectors, $\ket{\psi_{\x}} \otimes  \cdots \otimes \ket{\psi_{\x}}$. Using only two copies of each state allows the classifier to reach a $100\%$ success rate on classes $2$ and $3$. This trick furthermore allows us to consider datasets which are not based on angles, such as the concentric circles in fig. \ref{Fig:kerneltrick}, where the feature map again leads to perfect classification.\\  
The results from experiment and simulations suggest that the distance-based classifier realised by the interference circuit presents an interesting toy model with various potential extensions. Such a model can help to analyse quantum machine learning on real quantum processors, and create models for pattern recognition that are inspired by the strengths of quantum computing. Future work could aim at amending the circuit to realise different kernel functions that allow for more localised measures in order to increase the power and flexibility of the classifier, as well as considering circuits that make more use of quantum resources such as entanglement. 

\acknowledgments
This work was supported by the South African Research Chair Initiative of the Department of Science and Technology and the National Research Foundation. We acknowledge use of the IBM Quantum Experience for this work. The views expressed are those of the authors and do not reflect the official policy or position of IBM or the IBM Quantum Experience team.

\clearpage
\appendix

\section{Estimating the prediction error with confidence intervals}

The prediction result in the quantum circuit is encoded in the probabilities $1-p=|\alpha|^2$ and $p = |\beta|^2$ to measure the ``class qubit'' in state $\ket{0}$ or $\ket{1}$ respectively. To get an estimator $\hat{p}$ of the true probability $p$, we run the algorithm $R$ times to collect a sample of outcomes $Q_1,...,Q_R$. This corresponds to sampling from a Bernoulli distribution of a binary random variable $Q$ with expectation value $E[Q] = p$ and variance $ \sigma = \sqrt{\frac{p(1-p)}{R}}$. If $\hat{p} > 0.5$, the result of the classification is $1$, while for $\hat{p} < 0.5$ the result is $-1$. We want to get an estimate of how the error $\epsilon$ decreases with the sample size $R$ within a sufficiently high confidence level.\\

A common approach in statistics is to compute a maximum error related to a confidence level $z$. A $z$-value of $2.58$ corresponds to a confidence level of $99\%$, which indicates the proportion of confidence intervals around the estimator constructed from different samples in which we expect to find the true value $p$. Frequently used is the Wald interval which is suited for cases of large $R$ and $p \approx 0.5$, which we expect for the classifier due to the broad kernel function. The estimator for $p$ is constructed as $\hat{p} = 1/R \sum_{r=1}^{R} Q_r$. The maximum error $E = \hat{p} - p$ can be determined as 
$$E = z \sigma = z \sqrt{\frac{p(1-p)}{R}}. $$
This is maximised for $p = 0.5$, so that we can assume the overall error of our estimation $\epsilon$ to be at most $\frac{z}{2\sqrt{R}}$ with a confidence level of $z$. In other words, the number of times we have to repeat the classification algorithm (including state preparation) grows with $\mathcal{O}(\epsilon^{-2})$.\\

\newpage

For small sample sizes $R$, or $p$ close to either zero or one this estimation can fail severely \cite{brown01}. A more refined alternative is the Wilson score with the following estimator for $p$,
$$ \frac{1}{1+ \frac{z^2}{R}} \left( \hat{p} + \frac{z^2}{2R}\right),$$
and the maximum error
$$ \frac{z}{1+ \frac{z^2}{R}} \left( \frac{\hat{p}(1-\hat{p})}{R} + \frac{z^2}{4R^2}\right)^{\frac{1}{2}}. $$
Again this is maximised for $\hat{p} = 0.5$ and with a confidence level $z$ we can state that the overall error of our estimation is bounded by
$$ \epsilon \leq \sqrt{z^2\frac{R+z^2}{4R^2}}.$$
The more refined estimation therefore allows us to predict a runtime that grows only with $\mathcal{O}(\epsilon^{-1})$ (see also \cite{sweke16}).\\

\section{Details on the experiment with the IBM Quantum Experience}

\begin{center}
\begin{figure}[h!]
\centering
\includegraphics[scale=1]{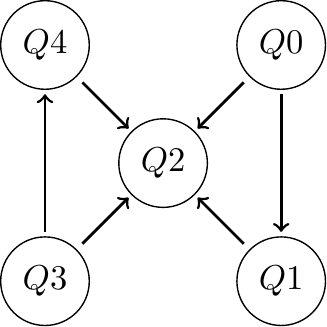}
\caption{\label{Fig:ibm_arrangement} Schematic drawing illustrating the qubit arrangement and the cross-resonance interactions on the IBM quantum processor chip.}
\end{figure}
\end{center}

At the time of writing IBM's quantum processor implements single-qubit gates within $83 \mathrm{ns}$
and a CNOT gate within maximally $483 \mathrm{ns}$ \cite{ibmgatetimes}. Figure \ref{Fig:ibm_arrangement} is a schematic drawing showing the cross-resonance interactions between and the arrangement of the five qubits in the IBM quantum computer.
According to the device calibration results from IBM \cite{ibm_calibration}, the amplitude damping times of the five qubits currently range from $46 \mathrm{ms}$ to $61.5 \mathrm{ms}$ and the phase damping times range from $38.6\mathrm{ms}$ to $91.9 \mathrm{ms}$. In the current hardware setup, the maximum single-qubit error and single-qubit measurement error are $3.3\times10^{-3}$ and $6.4\times10^{-2}$ respectively. As a result of these qubit errors and decoherence times, the system currently allows for 80 quantum operations per qubit thus enabling 79 quantum gates and one measurement.

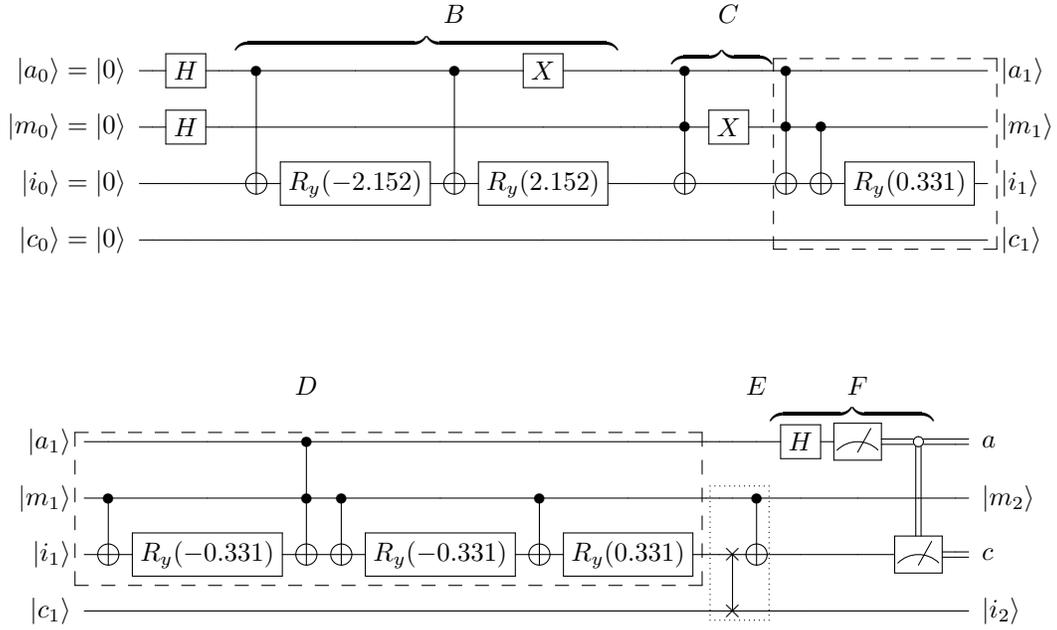
\begin{figure*}

$$
\Qcircuit @C=.5em @R=0.5em @!R {
& & & & & & & & & & & &  & B &  & &  & & &  & C &  & &  &  & \\
& & & & & &\lstick{\ket{a_0}=\ket{0}} & \qw & \gate{H} & \qw & \qw & \ctrl{2}  & \qw & \ctrl{2} &  \gate{X}  & \qw & \qw & \qw & \qw & \ctrl{1} & \qw & \qw & \ctrl{1} & \qw & \qw & \rstick{\ket{a_1}} \qw \\
& & & & & & \lstick{\ket{m_0}=\ket{0}} & \qw & \gate{H} & \qw & \qw &\qw & \qw & \qw & \qw & \qw & \qw & \qw & \qw  & \ctrl{1} & \gate{X} & \qw & \ctrl{1} & \ctrl{1} & \qw & \rstick{\ket{m_1}} \qw \\
& & & & & & \lstick{\ket{i_0}=\ket{0}} & \qw & \qw & \qw & \qw & \targ & \gate{R_y(-2.152)} & \targ & \gate{R_y(2.152)} & \qw  & \qw & \qw & \qw  & \targ & \qw & \qw & \targ & \targ & \gate{R_y(0.331)} & \rstick{\ket{i_1}} \qw\\
& & & & & & \lstick{\ket{c_0}=\ket{0}} & \qw & \qw & \qw & \qw & \qw & \qw & \qw & \qw & \qw & \qw & \qw & \qw & \qw & \qw & \qw & \qw & \qw & \qw & \rstick{\ket{c_1}} \qw \gategroup{2}{12}{4}{15}{.7em}{^\}} \gategroup{2}{20}{5}{22}{.7em}{^\}} \gategroup{2}{23}{5}{26}{.7em}{--}
}
$$\vspace{1em}

$$
\Qcircuit @C=.5em @R=0.5em @!R {
& & & D & & & & & & & & E & & F & &  & &  \\
\lstick{\ket{a_1}} & \qw & \qw & \ctrl{1} & \qw & \qw & \qw & \qw & \qw & \qw & \qw & \qw & \gate{H} & \meter & \cctrlo{2} & \cw & \rstick{a} \cw  \\
\lstick{\ket{m_1}} & \ctrl{1} & \qw & \ctrl{1} & \ctrl{1} & \qw & \ctrl{1} & \qw & \qw & \qw & \qw &  \ctrl{1} & \qw  & \qw & \qw & \qw  & \rstick{\ket{m_2}} \qw \\
\lstick{\ket{i_1}} & \targ & \gate{R_y(-0.331)} & \targ & \targ & \gate{R_y(-0.331)} & \targ & \gate{R_y(0.331)} & \qw & \qw & \qswap & \targ & \qw  & \qw & \meter & \cw & \rstick{c} \cw\\
\lstick{\ket{c_1}} & \qw & \qw & \qw & \qw & \qw  & \qw & \qw & \qw & \qw & \qswap \qwx & \qw & \qw & \qw & \qw & \qw & \rstick{\ket{i_2}} \qw \gategroup{2}{1}{4}{8}{.7em}{--}
\gategroup{3}{10}{5}{12}{.7em}{.} \gategroup{2}{13}{5}{15}{.7em}{^\}}\\
& & & & & & & & & & & & & & & & &  \\
}
$$\vspace{-1em}

\caption{\label{Fig:2D_full_circuit} Quantum circuit with decomposed controlled rotations implementing the distance-based classifier using the two training vectors $\x^0$ and $\x^1$ and the input vector $\tilde \x'$ from the rescaled and normalised Iris flower dataset. First the input vector is entangled with the ground state of the ancilla (Step B), then the training vector $\x^0$ is entangled with the excited state of the ancilla and the ground state of the index qubit (Step C) followed by entangling the training vector $\x^1$ with the excited state of the ancilla and the index qubit (Step D). Finally the data and class qubits are swapped and the class qubit is flipped conditioned on the index qubit being $\ket{1}$ (Step E). In step F after the Hadamard gate the ancilla is measured followed by a measurement of the class qubit when the ancilla was found to be in the $\ket{0}$ state.}
\end{figure*}

\begin{figure*}[t]
$$
\Qcircuit @C=1.3em @R=2em @!R {
& \qswap  & \qw & & & \qw & \ctrl{1} & \gate{H} & \ctrl{1} & \gate{H} & \ctrl{1} &  \qw \\
& \qswap \qwx & \qw & \push{\rule{.3em}{0em}=\rule{.3em}{0em}} & &
\qw & \targ & \gate{H} & \targ & \gate{H} & \targ &  \qw
}
$$\caption{\label{Fig:swap_decomp} Quantum circuit implementing a SWAP gate with four Hadamard and three CNOT gates.}
\end{figure*}
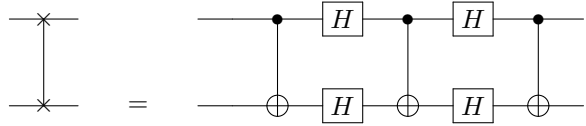

\vfill\eject
The full quantum circuit implementing the classification of input vector $\tilde \x'$ is given in Figure \ref{Fig:2D_full_circuit} where $\ket{a}$,$\ket{m}$,$\ket{i}$ and $\ket{c}$ stand for ancilla, index, data and class qubit respectively. Note, that the IBM Quantum Experience (IBMQE) does not provide all-to-all connected CNOT gates which follows directly from the qubit arrangement shown in Figure \ref{Fig:ibm_arrangement}. From Figure \ref{Fig:2D_full_circuit} it follows that the data qubit $\ket{i}$ is the most frequently used target qubit for controlled operations. The third qubit ($Q2$ in Figure \ref{Fig:ibm_arrangement}) is the only qubit which can be connected to all other qubits by means of CNOT gates and is, thus, chosen to be the data qubit. To flip the class label for the training vector $\x^1$ a CNOT gate needs to be applied to the class qubit controlled by the index qubit (Step D in Figure \ref{Fig:2D_full_circuit}). The available CNOT connectivity of the IBM quantum computer requires prior swapping of the data and class qubits as indicated in Step D in Figure \ref{Fig:2D_full_circuit}. On the IBMQE a SWAP gate can be implemented with the quantum circuit shown in Figure \ref{Fig:swap_decomp}.

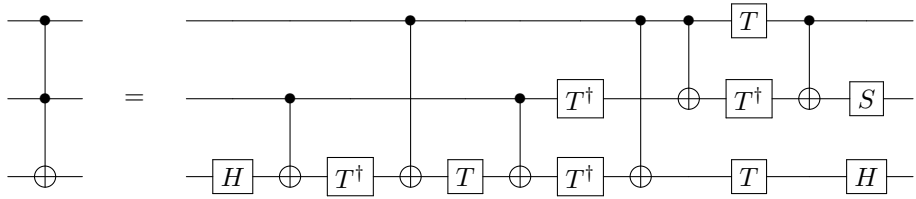
\begin{figure*}
$$
\Qcircuit @C=1em @R=1.5em @!R {
& \ctrl{1} & \qw & & & \qw & \qw & \qw & \ctrl{2} & \qw & \qw & \qw & \ctrl{2} & \ctrl{1} & \gate{T} & \ctrl{1} & \qw & \qw \\
& \ctrl{1} & \qw &
\push{\rule{.3em}{0em}=\rule{.3em}{0em}} & &
\qw & \ctrl{1} & \qw & \qw & \qw & \ctrl{1} & \gate{T^\dagger} & \qw & \targ & \gate{T^\dagger} & \targ & \gate{S} &  \qw\\
& \targ & \qw & & & \gate{H} & \targ & \gate{T^\dag} & \targ & \gate{T} & \targ & \gate{T^\dagger} & \targ & \qw  & \gate{T} & \qw  & \gate{H} & \qw
}
$$\caption{\label{Fig:toffoli_decomp} Quantum circuit implementing a Toffoli gate with ten single-qubit gates (T-depth 4) and six CNOT gates.}
\end{figure*}

\newpage
The entire quantum state preparation routine outlined in Figure \ref{Fig:2D_full_circuit} requires the use of three Toffoli gates in Steps B and C. Toffoli gates are not directly supported by the IBM quantum hardware and, thus, need to be decomposed as shown in Figure \ref{Fig:toffoli_decomp}. There are many known ways of decomposing a Toffoli gate but we specifically chose this decomposition since it integrates very well with IBM's CNOT connectivity and has a relatively low T-depth of four.

For the classification of the second input vector $\tilde \x''$ the overall quantum circuit shown in Figure \ref{Fig:2D_full_circuit} remains the same. To load $\tilde \x''$ instead of $\tilde \x'$ only the rotations in Step A need to be changed to $R_y(-2.152)$ and $R_y(2.152)$ instead of $R_y(-1.518)$ and $ R_y(1.518)$.

Table \ref{tab:raw_results} shows the obtained occurence counts for all four-qubit quantum states after 8192 runs in the classification of input vectors $\tilde \x'$ and $\tilde \x''$. 

\begin{table*}
\def\arraystretch{1.4} \renewcommand{\tabcolsep}{3.4pt}
\begin{tabular}{lcccccccccccccccc}

\hline
\hline
 &$\ket{0}$&$\ket{1}$&$\ket{2}$&$\ket{3}$&$\ket{4}$&$\ket{5}$&$\ket{6}$&$\ket{7}$&$\ket{8}$&$\ket{9}$&$\ket{10}$&$\ket{11}$&$\ket{12}$&$\ket{13}$&$\ket{14}$&$\ket{15}$\\\hline
$\x'$ &773 & 1400 & 223 & 172 & 210&205&1013&578&823&1175&117&113&95&114&476&705\\
$\x''$&948&1117&166&145&155&128&680&626&1139&1136&131&122&127&149&694&729\\
\hline
\hline
\end{tabular}
\caption{Raw experimental results for the classification of the input vectors $\tilde \x'$ and $\tilde \x''$. The table shows the occurence counts for each four-qubit quantum state, $\ket{0000} \rightarrow \ket{0}, \ket{0001} \rightarrow \ket{1}...$,  in 8192 runs.}
\label{tab:raw_results}
\end{table*}

\end{document}